\documentclass[runningheads,a4paper]{llncs}

\usepackage{amssymb}
\usepackage{graphicx}

\usepackage{url}
\newcommand{\keywords}[1]{\par\addvspace\baselineskip
\noindent\keywordname\enspace\ignorespaces#1}

\begin{document}

\mainmatter  % start of an individual contribution

% first the title is needed
\title{Embedding Digital Signature into CSV Files \\Using Data Hiding}

% a short form should be given in case it is too long for the running head
\titlerunning{Digital Signature for CSV}

% the name(s) of the author(s) follow(s) next
%
% NB: Chinese authors should write their first names(s) in front of
% their surnames. This ensures that the names appear correctly in
% the running heads and the author index.
%
\author{Akinori Ito}
\authorrunning{Akinori Ito}

% the affiliations are given next; don't give your e-mail address
% unless you accept that it will be published
\institute{Graduate School of Engineering, Tohoku University\\
6-6-5 Aramaki aza Aoba, Aoba-ku, Sendai, 980-8579 Japan\\
\path|aito.spcom@tohoku.ac.jp|
}

%
% NB: a more complex sample for affiliations and the mapping to the
% corresponding authors can be found in the file "llncs.dem"
% (search for the string "\mainmatter" where a contribution starts).
% "llncs.dem" accompanies the document class "llncs.cls".
%

\toctitle{Lecture Notes in Computer Science}
\tocauthor{Authors' Instructions}
\maketitle

\begin{abstract}
Open data is an important basis for open science and evidence-based policymaking. Governments of many countries disclose government-related statistics as open data. Some of these data are provided as CSV files. However, since CSV files are plain texts, we cannot ensure the integrity of a downloaded CSV file. A popular way to prove the data's integrity is a digital signature; however, it is difficult to embed a signature into a CSV file. This paper proposes a method for embedding a digital signature into a CSV file using a data hiding technique. The proposed method exploits a redundancy of the CSV format related to the use of double quotes. The experiment revealed we could embed a 512-bit signature into actual open data CSV files.

\end{abstract}

\section{Introduction}

Open data is defined as data that is freely accessible, usable, reusable, and redistributable by anyone, subject only to the requirement of attribution and share-alike principles. This concept is integral to the broader movement towards openness and sharing, encompassing open content, knowledge, and resources.

The primary characteristics of open data include:
\begin{itemize}
\item \textbf{Accessibility}: Data should be available to all individuals without discrimination \cite{machova2018usability}.
\item \textbf{Reusability}: Data should be provided conveniently and modifiable, facilitating easy reuse and repurposing \cite{zuiderwijk2019sharing}.
\item \textbf{Redistribution}: Individuals must be free to redistribute the original or modified versions of the data.
\item \textbf{Attribution}: While the data can be used freely, certain open data licenses require users to credit the source of the data.
\end{itemize}

Another aspect of open data is the FAIR principle \cite{wilkinson2016fair,jacobsen2020fair}, where FAIR stands for Findable, Accessible, Interoperable, and Reusable. Here, ``interoperability'' refers to the ability of data to be integrated with other data and to work with various applications or workflows for analysis, storage, and processing. 

Open data is frequently associated with government data \cite{Okamoto2016WhatIB}, as numerous governments globally have committed to making their data openly available to enhance transparency, civic engagement, and innovation. The application of open data has been investigated in various contexts, demonstrating its potential to improve data provision and policymaking \cite{Loenen2017LogFA}, contribute to sustainable development \cite{fasli2023open}, enhance transversal skills and global citizenship in education \cite{Atenas2015OpenDA}, and support open science \cite{Numajiri2024}.

Several studies have examined the data formats utilized in open data initiatives. Oliveira et al. \cite{Oliveira2016OpenGD} identified that Brazilian OGD portals predominantly use CSV, among other formats. Washington and Morar \cite{Washington2017OpenGD} discussed the impact of file formats on collaboration potential, noting that data.gov formats have limited potential but are accessible to users with diverse skills. 
%Krataithong et al. \cite{Krataithong2016SemiautomaticFF} proposed a framework for converting tabular data to RDF format, enhancing data querying and integration. 
Yi \cite{Yi2019ExploringTQ} emphasized the importance of data quality, particularly regarding data completeness and machine-readability, and suggested guidelines for data format selection and data completeness enhancement.

The Comma-Separated Values (CSV) \cite{shafranovich2005common} have been widely used as an interoperable data format and method for enhancing its utility. 
A distinct feature of the CSV format compared to other data formats, such as JSON or XML, is its simplicity and ease of use for tabular data representation. CSV files store data in plain text, where each line corresponds to a data record, and each record consists of fields separated by commas. This straightforward structure makes CSV files highly accessible and easy to manipulate using basic text editors and simple programming scripts.
For this reason, many open data files are provided as CSV files \cite{Mitloehner2016}.

The CSV format is a simple, plain text format ideal for storing and exchanging tabular data due to its lightweight nature and broad compatibility across platforms. CSV files are limited to raw data storage without additional functionalities, unlike the more complex Excel format, which supports advanced features like formatting, formulas, and data visualization. This simplicity ensures smaller file sizes and faster processing, making CSV suitable for large datasets. In contrast, Excel's robust data manipulation capabilities make it better suited for comprehensive data analysis and visualization tasks. However, it may suffer from larger file sizes and compatibility issues outside of Microsoft Excel.

Because CSV files are interoperable, several projects have tried to handle them for further data use.
Alkarkoukly et al. \cite{Alkarkoukly2023} present a reference implementation for transforming CSV data into FHIR resources using open-source tools, addressing interoperability challenges in healthcare. Mahmud et al. \cite{Mahmud2018} propose an approach to generate annotated tables from CSV files, improving semantic structure and reusability. Christodoulakis et al. \cite{Christodoulakis2020} introduce Pytheas, a method for automatically discovering tables within CSV files, outperforming existing approaches in precision and recall. Mahmud et al. \cite{Mahmud2019} developed a semantic approach to convert CSV data into RDF format and publish it as Linked Open Data on the web, following W3C recommendations. These studies collectively demonstrate the versatility of CSV files and propose solutions to enhance their interoperability, semantic richness, and integration with web technologies, addressing challenges in various domains, including healthcare, open government data, and semantic web applications.

A problem with the CSV file format is that it does not have a mechanism to ensure data security. Since the open data provided by governments are the basis of public research and surveys, distributing tampered data can considerably threaten society. The digital signature \cite{lax2015digital} is a framework to ensure authenticity and integrity, commonly used for document workflow \cite{pop2010digital}. When the data formats for Microsoft Excel (such as xls or xlsx) are used, we can sign the file using the digital signature technique \cite{MSsignature}. 
However, since the CSV format is a simple text file, inserting a digital signature into a CSV file is impossible.
Therefore, this paper proposes a method to embed a digital signature into a CSV file using a data hiding technique. 

\section{Previous Work}
\subsection{CSV file format}
CSV is a file format widely used for tabular data. It is a plain text, where a line makes a record and items in a record are separated by commas.
Fig.\ref{fig:csv} shows an example of a CSV file where the data has one header and two records, each with four fields. Each field, such as \texttt{H1} or \texttt{Hello}, may be enclosed by double quotes. 

\begin{figure}[tb]
\centering
\begin{tabular}{|l|}
\hline
\tt H1,H2,H3,N1 \\
\tt "Hello","my","world",3.0 \\
\tt "Nice","to","meet",8.0 \\
\hline
\end{tabular}
\caption{An example of a CSV file.}
\label{fig:csv}
\end{figure}

In actual open data, the format of CSV files is not necessarily consistent. According to Mitl\"oner et al. \cite{Mitloehner2016}, only 88\% of open data use commas as field separators and 10\% use semicolons. Therefore, there was an attempt to make a standard of the CSV format as RFC \cite{shafranovich2005common}. 
Fig.\ref{fig:RFC4180} shows the syntax description of a CSV file proposed in RFC4180 written in ABNF \cite{ABNF}. According to the standard (RFC 4180), a CSV file is composed of one or more records delimited by CRLFs. The file may have a header line, but there is no way to determine whether or not the first line is the header line. 
A record (and a header) has multiple fields, where a field is either an escaped or a non-escaped string. An escaped string is a string enclosed by double quotes, where the string inside may contain commas, CRs, LFs, and double quotes (a double quote is escaped by one more double quote). A non-escaped string can contain any character besides commas, double quotes, CRs, and LFs.

\begin{figure}[tb]
\centering
\begin{tabular}{|l|}
\hline
file = [header CRLF] record *(CRLF record) [CRLF] \\
header = name *(COMMA name) \\
record = field *(COMMA field) \\
name = field \\
field = (escaped / non-escaped) \\
escaped = DQ *(TEXTDATA / COMMA / CR / LF / DQ DQ) DQ \\
non-escaped = *TEXTDATA \\
COMMA = \%x2C \\
CR = \%x0D \\
DQ = \%x22 \\
LF = \%x0A \\
CRLF = CR LF \\
TEXTDATA = \%x20-21 / \%x23-2B / \%x2D-7E \\
\hline
\end{tabular}
\caption{The syntax description of the CSV file format \cite{shafranovich2005common}.}
\label{fig:RFC4180}
\end{figure}

\subsection{Digital signature}
A digital signature is a robust cryptographic mechanism that guarantees the authenticity and integrity of a digital document \cite{rivest1978method,lax2015digital} . It harnesses a blend of mathematical algorithms and cryptographic keys to achieve a high level of security, providing reassurance in the digital realm.

The core principle behind a digital signature involves using a public-key cryptography system \cite{rivest1978method}. The signer possesses a private key, which is kept confidential, and a corresponding public key, which can be freely distributed. 
When a document is signed, a mathematical hash function is applied, generating a unique digital fingerprint. 
This fingerprint (a hash value) is then encrypted with the signer's private key, creating the digital signature.

The recipient of the signed document can verify its authenticity using the signer's public key. The recipient obtains the original hash value by decrypting the signature with the public key. They can then independently calculate the hash of the received document and compare it to the decrypted value. If both values match, it confirms that the document originated from the claimed signer and has not been tampered with since the signature was applied.

Digital signature streamlines workflows by enabling secure electronic document signing and verification, eliminating the need for physical documents and manual processing \cite{pop2010digital}. It is also used to maintain the authority and integrity of open data. Wong et al. proposed a system architecture for signing open data \cite{wong2015architecture}. Their proposal includes a public key infrastructure and HTTP servers that provide the data. However, they did not discuss how the signature is attached to the data file; it may be assumed that the provided data is in a format that can involve a signature, such as xlsx.

Unfortunately, a CSV file cannot include a signature because it is a plain text file with no inner structure. Therefore, even if we calculate a signature for a CSV file, we need to deliver it separately, which may reduce the value of the signature.

\section{Proposed Method}
\subsection{Data hiding in a CSV file}
The proposed method embeds arbitrary binary data into a CSV file. To do that, we exploit the redundancy of quoting a field. As shown in Fig.\ref{fig:RFC4180}, a field can be either an escaped string or a non-escaped string when the string does not involve special characters (double quote, comma, CR, and LF).

Consider a CSV file has $N$ rows and $M$ columns\footnote{The data in a CSV file do not need to be in a single table. However, we assume that it contains a rectangular table here for simplicity; the following discussion applies to non-rectangular data.}. 
\begin{equation}
X=\left[\begin{array}{ccc}
x_{11}&\cdots&x_{1M}\\
\vdots & & \vdots \\
x_{N1}&\cdots&x_{NM}\\
\end{array}\right]
\end{equation}
Here, $x_{ij}$ is a character string, including a null string, without enclosing double quotes.
Then we regard all the fields as a sequence, $x(1), \ldots, x(K)$, where $x_{ij}=x((i-1)M+j)$ and $K=NM$.  We consider a function to determine whether a field needs to be escaped:
\begin{equation}
\mathrm{noesc}(x) = \left\{
\begin{array}{ll}
0 & \quad\mathrm{if}~x~ \textrm{contains any of comma, double quote, CR, or LF} \\
1 & \quad\mathrm{otherwise}
\end{array}
\right.
\end{equation}
If $\mathrm{noesc}(x)=0$, it means that the string $x$ should be enclosed by double quotes. Otherwise, $x$ may be enclosed by double quotes or it may be used without double quotes. Therefore, the fields where $\mathrm{noesc}(x)=1$ can be used for a one-bit payload. The total payload can be calculated as $P(K)$, where
\begin{equation}
P(k)=\sum_{i=1}^k \mathrm{noesc}(x(i)).
\end{equation}
$P(k)$ is the number of bits we can embed into $x(1),\ldots,x(k).$

Next, we define two functions to calculate the string representation of the data. 
The first is to put the double quotes in only the needed fields.
\begin{equation}
J_\mathrm{simp}(X)=\mathop{\mathrm{Con}_\mathrm{\tiny CRLF}}_{i=1}^N \left(
\mathop{\mathrm{Con}_,}_{j=1}^M \mathrm{esc}_0(x_{ij}) \right)
\end{equation}
Here, $\mathrm{Con}_\alpha s_i$ means concatenation of strings $s_1,s_2,\ldots$ using delimiter $\alpha$, as follows.
\begin{equation}
\mathop{\mathrm{Con}_\alpha}_{i=1}^M s_i = s_1 \circ \alpha \circ s_2 \circ \alpha \circ \cdots \circ \alpha \circ s_M,
\end{equation}
where $\circ$ means the string concatenation operator. $\mathrm{esc}_0$ is an escape function defined as follows.
\begin{equation}
\mathrm{esc}_0(x)=\left\{\begin{array}{cc}
\texttt{"}\circ x \circ \texttt{"} & \textrm{if noesc}(x)=0 \\
x & \textrm{if noesc}(x)=1 \\
\end{array}
\right.
\end{equation}

The second function is to embed data into the CSV string. First, we consider the data to be embedded as follows.
\begin{equation}
\mathbf{b}=b_1,\ldots,b_{P(K)},\quad b_k\in\{0,1\}.
\end{equation}
Then, the function is defined as follows.
\begin{eqnarray}
J_\mathrm{emb}(X,\mathbf{b})&=&\mathop{\mathrm{Con}_\mathrm{\tiny CRLF}}_{i=1}^N \left(
\mathop{\mathrm{Con}_,}_{j=1}^M \mathrm{esc}_1(x_{ij},b_k) \right)\\
&&\mathrm{where}~k=P((i-1)M+j) \nonumber
\end{eqnarray}
\begin{equation}
\mathrm{esc}_1(x,b)=\left\{
\begin{array}{ll}
\texttt{"}\circ x \circ \texttt{"} &\quad \mathrm{if}~\mathrm{noesc}(x)=0~\mathrm{or}~b=1 \\
x &\quad\mathrm{otherwise}
\end{array}
\right.
\end{equation}
The function $\mathrm{esc}_1(x)$ encloses $x$ with double quotes when $x$ involves special characters or the bit 1 is embedded.
Fig. \ref{fig:csv_embed} shows an example of a CSV file with a hidden message. This CSV file has nine fields (three rows, three columns), but the first field of the second row (\texttt{"Hello,world"}) cannot be used as a payload because it contains a comma. 

\begin{figure}[tb]
\centering
\begin{tabular}{|l|}
\hline
\tt "H1",H2,"N1" \\
\tt "Hello,world","green",3.0 \\
\tt Nice to meet you,"apple","8.0" \\
\hline
\end{tabular}
\caption{An example of a CSV file with embedded information. The payload is 8 bits, and the message 10110011 is embedded.}
\label{fig:csv_embed}
\end{figure}

\subsection{Extraction of contents and messages}
Consider a CSV file in which each field may or may not be enclosed by double quotes. Then, consider a matrix made by splitting the input CSV file using CRLFs and commas.
\begin{equation}
Y=\left[\begin{array}{ccc}
y_{11}&\cdots&y_{1M}\\
\vdots & & \vdots \\
y_{N1}&\cdots&y_{NM}\\
\end{array}\right]
\end{equation}
Here, $y_{ij}$ may or may not be enclosed by double quotes. For example, when the input CSV file contains 
\begin{center}
\texttt{"H1",H1,"N1"} CRLF \texttt{"Hello,world","green",3.0},
\end{center}
then the comma-split data is
\begin{equation}
Y=\left[\begin{array}{ccc}
\texttt{"H1"} & \texttt{H1} & \texttt{"N1"} \\
\texttt{"Hello,world"} & \texttt{"green"} & \texttt{3.0} \\
\end{array}\right].
\end{equation}

Then, we define two functions, $\mathrm{strip}(y)$ and $\mathrm{extract}(y)$, as follows.
\begin{equation}
\mathrm{strip}(y) = \left\{
\begin{array}{ll}
x &\quad \mathrm{if}~y=\texttt{"}\circ x\circ \texttt{"}\\
y &\quad \mathrm{otherwise}
\end{array}
\right.
\end{equation}
%\begin{equation}
%\mathrm{strip}(y) = \left\{
%\begin{array}{ll}
%\mathrm{st}(y) &\quad \mathrm{if}~\mathrm{noesc}(\mathrm{st}(y))=1\\
%y &\quad \mathrm{otherwise}
%\end{array}
%\right.
%\end{equation}
\begin{equation}
\mathrm{extract}(y) = \left\{
\begin{array}{ll}
0 &\quad \textrm{if}~\mathrm{strip}(y)=y\\
1 &\quad \textrm{else if}~\mathrm{noesc}(\mathrm{strip}(y))=1\\
\epsilon & \quad \mathrm{otherwise}
\end{array}
\right.
\end{equation}
Here, $\mathrm{strip}(y)$ is a function that trims the enclosing double quotes. The function $\mathrm{extract}(y)$ is a function to extract the hidden bit from a field, which returns one of 0, 1, and $\epsilon$. The value $\epsilon$ means that the field does not contain a hidden bit.

Using $\mathrm{extract}(y)$, we can extract the hidden message from $Y$ by concatenating all of $\mathrm{extract}(y(1)),\ldots,\mathrm{extract}(y(K))$ and removing all $\epsilon$ from the sequence. We denote this sequence as $\mathrm{extract}(Y).$

\subsection{Insertion and validation of digital signature}

First, we consider stripping all unnecessary double quotes from the CSV file:
\begin{equation}
\mathrm{strip}(Y)=\left[\begin{array}{ccc}
\mathrm{strip}(y_{11})&\cdots&\mathrm{strip}(y_{1M})\\
\vdots & & \vdots \\
\mathrm{strip}(y_{N1})&\cdots&\mathrm{strip}(y_{NM})\\
\end{array}\right]
\end{equation}
Applying strip to $Y$ removes all enclosing double quotes for all fields. 

When calculating a signature, we need a pair of the public and private keys, denoted as $\kappa_\mathrm{\scriptsize pub}$ and $\kappa_\mathrm{\scriptsize priv}$, respectively.

Then, we insert a signature and validate it as follows. Let a hash function be $H(s)$ that returns a fixed-sized bit sequence, where $s$ is a string. 
Let an encryption function be $E(\mathbf{b},\kappa)$, which receives a bit sequence $\mathbf{b}$ and a key $\kappa$, then returns a bit sequence with the same length as $\mathbf{b}.$ When we have data $Y$ from a CSV file, we calculate the signature $\mathbf{s}$ as
\begin{equation}
\mathbf{s}=E(H(J_\mathrm{simp}(\mathrm{strip}(Y))),\kappa_\mathrm{priv})).
\end{equation}
Then, we generate the embedded CSV file as
\begin{equation}
CSV=J_\mathrm{emb}(\mathrm{strip}(Y),\mathbf{s})
\end{equation}

Next, when we receive a signed CSV file, we calculate a matrix $Y'$ by splitting the CSV files by CRLFs and commas. Then, we extract the embedded signature as
\begin{equation}
\mathbf{s}'=\mathrm{extract}(Y').
\end{equation}
Now we decrypt $\mathbf{s}'$ and compare it to the hash extracted from the stripped CSV file. If
\begin{equation}
E(\mathbf{s}',\kappa_\mathrm{pub})=H(J_\mathrm{simp}(\mathrm{strip}(Y')),
\end{equation}
then the CSV file is validated.

\section{Experiment}

I carried out a proof-of-concept experiment. The program was implemented using R 4.4.0 with openssl package. 
Ten CSV files were downloaded from e-gov.go.jp and used for the experiment. All data were open data provided by the Japanese government.
Table \ref{tab:data} shows the data. Note that we used data with a payload larger than 512 bits because the signature size was 512 bits.
We used RSA key pair and 512-bit SHA1 hash. 
We zero-padded the embedding bits since all payloads were larger than the signature size.

\begin{table}[tb]
\caption{Data used for the experiment. All data were downloaded from e-gov.go.jp.}
\label{tab:data}
\centering
\begin{tabular}{lrr}
\hline
Filename & Filesize (byte) & Payload (bit) \\
\hline
081\_AL\_01s\_2009.csv & 1869 & 618 \\
151\_AB\_01s\_2006.csv & 31081 & 5598 \\
181\_AB\_01s\_2009.csv & 31911 & 5598 \\
mc010000.csv & 5864 & 1063 \\
mc070000.csv & 4781 & 821 \\
mc110000.csv & 12708 & 1764 \\
s4-7.csv & 7024 & 880 \\
tk9003.csv & 2164 & 581 \\
tk9005.csv & 3875 & 1181 \\
tk9012.csv & 3924 & 696 \\
\hline
\end{tabular}
\end{table}

First, I confirmed whether the embedded CSV files were adequately validated. As a result, all files were validated using the public key.

Next, I checked that the tampered files were detected. To do that, I used LibreOffice to open the CSV files and saved them in CSV format. Although I did not change the contents of the files, loading and saving the files removed the double quotes. As a result, all files were not validated using the public key.

\section{Conclusion}
This paper proposed a new method for embedding digital signatures into CSV files. The proposed method uses a redundancy: We may or may not enclose the content of a field with double quotes as long as the content does not contain special characters.

There are still several problems in this framework, including the following points:
\begin{itemize}
\item 
Since the payload is limited, embedding a stronger signature, such as 1024 or 2048 bits, could be difficult.
\item Because of the same reason, it could be difficult to embed other metadata, such as the signer, timestamp, and digital certificate.
\end{itemize}
These problems should be solved in a future work.

\bibliographystyle{splncs03}
\bibliography{reference}

\end{document}